\documentclass[reprint,secnumarabic, amssymb, amsmath, aps,prb]{revtex4-1}

\usepackage{docs,graphicx,bm}
\usepackage{systeme,braket}
\usepackage[colorlinks=true,linkcolor=blue]{hyperref}%
\expandafter\ifx\csname package@font\endcsname\relax\else
\expandafter\expandafter
\expandafter\usepackage
\expandafter\expandafter
\expandafter{\csname package@font\endcsname}%
\fi
\begin{document}

\title{Asymmetry-driven plasmon instabilities in confined hydrodynamic electron flows}

\author{Aleksandr S. Petrov}%
\email{aleksandr.petrov@phystech.edu}
\affiliation{Laboratory of 2D Materials' Optoelectronics, Moscow Institute of Physics and Technology, Dolgoprudny 141700,	Russia}%

\author{Dmitry Svintsov}%
\affiliation{Laboratory of 2D Materials' Optoelectronics, Moscow Institute of Physics and Technology, Dolgoprudny 141700,	Russia}%

\date{18 December 2017}%
\revised{20 December 2017}%

\begin{abstract}
Direct current in confined two-dimensional (2d) electron systems can become unstable with respect to the excitation of plasmons. Numerous experiments and simulations hint that structural asymmetry somehow promotes plasmon generation, but a constitutive relation between asymmetry and instability has been missing. We provide such relation in the present paper and show that bounded perfect 2d electron fluids in asymmetric structures are unstable under arbitrarily weak drive currents. To this end, we develop a perturbation theory for hydrodynamic plasmons and evaluate corrections to their eigenfrequency induced by carrier drift, scattering, and viscosity. We show that plasmon gain continuously increases with degree of plasmon mode asymmetry until it surrenders to viscous dissipation that also benefits from asymmetry. The developed formalism allows us to put a lower bound on the instability threshold current, which corresponds to the Reynolds number $R_{\min} = 2\sqrt{3}$ for one-dimensional plasmons in 2d channel under constant voltage bias.
\end{abstract}

\maketitle

\section{Introduction and outline}

Hydrodynamic transport has long been the experimental routine in gas and fluid mechanics yet unreachable for electrons in solids. The main reason was that the perquisite of hydrodynamics (HD), the dominance of electron-electron scattering over disorder scattering~\cite{gantmakher2012carrier}, could be fulfilled only in ultra-clean samples in a limited temperature window~\cite{deJong1995hydrodynamic}. With the advent of graphene-based heterostructures with low impurity density and high phonon energy~\cite{Mayorov2011graphenehBN}, the hydrodynamic window was considerably extended~\cite{narozhny2017hydrodynamic}. This enabled the observation of striking and counter-intuitive hydrodynamic phenomena, such as electronic whirlpools~\cite{bandurin2016negative}, higher-than-ballistic conduction~\cite{Levitov2017superballistic,kumar2017superballistic}, and breakdown of Wiedemann-Franz law~\cite{crossno2016observation}.

An intriguing yet weakly explored area of electron hydrodynamics is related to turbulence, i.e. instabilities of electron drift~\cite{Mendoza2011Preturbulent,Mendoza2017KelvinHelmholtz}. The first observations of hydrodynamic plasma instabilities~\cite{Kopylov1980instability} and turbulence~\cite{kopylov1987perioddoubling} in macroscopic solid state samples in 1980's remained almost unheeded. A strong effort in this field was triggered by the statement that HD electron flow in a confined gated two-dimensional electron system (2DES) is unstable with respect to the self-excitation of plasmons under certain boundary conditions, now known as Dyakonov-Shur (DS) instability~\cite{dyakonov1993shallow}. A similar situation occurs in wind music instruments when a steady air flow becomes unstable with respect to the excitation of sound~\cite{rayleigh1896theory}. An observable outcome of instability in confined 2DES is the electromagnetic emission due to radiative plasmon decay~\cite{chaplik1985absorption}. 

The physics beyond the original DS instability is the amplified reflection of downstream electron density perturbations from the drain contact~\cite{dyakonov1993shallow,sydoruk2012amplifying}, akin to the photon energy gain reflected from a moving object~\cite{zeldovich1972amplification}. Such amplification occurs only if the drain impedance much exceeds the impedance at source~\cite{cheremisin1999d}, the condition scarcely fulfilled in reality. Nevertheless, the electromagnetic emission from 2d field-effect transistors (FETs) was observed, evolving from noise-like spectra at liquid helium temperature in first experiments~\cite{knap2004terahertz} to gate-tunable resonant emission at $T=300$ K in advanced structures~\cite{ElFatimy2010algan}. 

The demonstration of ''electronic flute'' in spite of boundary condition problem stimulated dozens of instability proposals in related systems, including ungated~\cite{dyakonov2005current,Jena2009HydroConfined} and partly gated transistors~\cite{ryzhii2005transit,petrov2016plasma}, Corbino discs~\cite{sydoruk2010corbino}, and edges of 2DES~\cite{dyakonov2008boundary}. Both experiments and theories for numerous plasmonic structures hinted that structural asymmetry somehow promotes instability. However, the link between asymmetry and instability remained elusive so far, contrary to the well-established symmetry constraint for the inverse process, photodetection~\cite{sturman1992photovoltaic}.

In this paper, we provide a constitutive relation between asymmetry of confined 2DES-based plasmonic resonators and stability of their plasmon modes. To this end, we develop an analogue of quantum-mechanical perturbation theory for hydrodynamic equations in confined 2DES and express the growth/decay rates of plasmons through their field distributions in the absence of drift and dissipation. We show that the necessary condition for instability is the asymmetry of plasmon field: the mode gain turns to zero for even and odd field distributions, but is generally nonzero for modes without certain parity. This statement is similar to the presence of optical instabilities in relativistic systems with broken parity-time symmetry~\cite{Silveirinha2014ParityTime}. The asymmetry in confined 2DES can be induced by inequivalence of source and drain contacts (which is the case of famous DS instability), asymmetry of gating environment, non-uniform doping, or all of them. The gain in perfect electron fluid continuously increases with degree of mode asymmetry, and it is the electron viscosity that constraints this growth in highly asymmetric structures. We also pose the variational problems for determination of ''optimal plasmon modes'' that are most susceptible to instabilities, and find the lower bound for the instability threshold current.

\section{Perturbation theory for electron hydrodynamics}

Electrons in solid state are described by hydrodynamic equations if their mean free time with respect to e-e collisions $\tau_{ee}$ is much shorter than momentum relaxation time $\tau_p$ due to impurities and phonons and frequency of external field $\omega$. The confinement of a 2DES to a characteristic length $L$ leads to emergence of collective modes (plasma modes) with frequencies $\omega_p \sim [n_0 e^2 / m \varepsilon L]^{1/2}$, where $n_0$ is the electron density, $m$ is the effective mass, and $\varepsilon$ is the background dielectric constant. We are to study the stability of these modes under passage of direct current $j_0 = n_0 u_0$ without specifying the structure geometry and distribution of steady-state quantities.

The strategy of solution follows from the fact that the growth rate of asymmetry-related instabilities is a linear function of drive current~\cite{dyakonov1993shallow,sydoruk2010corbino,petrov2016plasma} in a nearly perfect electron fluid (i.e., in the absence of viscosity and momentum non-conserving scattering). Another observation is that direct current is manifested in convective terms in continuity and Navier-Stokes equations which can be well-isolated from ''non-drifting'' terms. Therefore, we are planning to construct the perturbation theory for frequencies of plasmons with respect to drift, viscosity, and momentum non-conserving collisions. The effects of gain and loss are expected to appear already in its first order.

The perturbation theory for HD equations of viscous charge-neutral incompressible fluid was developed by Joseph and Sattinger~\cite{joseph1972bifurcating}. It turned out to be quite complicated due to viscous dissipation in unperturbed equations; in other words, the unperturbed system was non-hermitian. The situation for charged confined fluid is simplified due to the presence of discrete plasma modes whose frequencies much exceed the corrections due to viscosity and drift. Hence, the ''unperturbed'' equations of motion can be chosen to be hermitian. The formal inequalities for perturbative treatment to be possible are $\{u_0/L,\tau^{-1}_p,\nu/L^2\}\ll \omega_{p}$, where $\nu$ is the kinematic viscosity. Taking the realistic parameters $u_0 \simeq 10^5$\,m/s, $L\simeq 1$ $\mu$m, $\tau_p \simeq 10^{-11}$ s$^{-1}$, and estimating the viscosity as $\nu\simeq v^2_0 \tau_{ee}/4\simeq 250$ cm$^2$/s~\cite{kumar2017superballistic}, we see that inequalities are fulfilled for $\omega_p/2\pi \simeq 1$ THz.

The set of hydrodynamic equations for one-dimensional motion of charged fluid has the form:
\begin{gather}
\label{eq-continuity-full}
\partial_t \mathcal N + \partial_x J = 0;\\
\label{eq-Euler-full}
\partial_t J + \partial_x \mathcal P = e \mathcal N \partial_x\varphi - J/\tau_{\rm p},
\end{gather}
where $t$ and $x$ stand for time and coordinate, respectively; $\mathcal N$ is the electron density, $J = \mathcal N \mathcal U$ is the current, $\mathcal U$ is the drift velocity, and $\mathcal P$ is the stress tensor of electron fluid (reduced to a scalar for one-dimensional motion):
\begin{equation}
    \mathcal P = \mathcal N \mathcal U^2 - \frac{\eta}{m}\partial_x \mathcal U,
\end{equation}
$\eta$ stands for dynamic viscosity. The set (\ref{eq-continuity-full}-\ref{eq-Euler-full}) is completed by the expression for electric potential $\varphi$ having the contributions from contacts $\varphi_{ext}(x)$ and self-consistent field:
\begin{equation}
    \label{eq-Poisson-full}
\varphi(x) = \varphi_{ext}(x) - e\int_0^L G(x,x^\prime)\mathcal N (x^\prime)\mathrm dx^\prime,
\end{equation}
here $G(x,x^\prime)$ is the Green's function of the electrostatic problem. The contribution $\varphi_{ext}(x)$ is fixed at contacts by the voltage source, $\varphi_{ext}(0) = 0$, $\varphi_{ext}(L) = V_D$, while the Green's function vanishes at these points.

The following mode stability analysis is based on linearization $\mathcal N(x,t) = n_0(x) + n(x)e^{-i\Omega t}$, $\mathcal U (x,t) = u_0(x) + u(x)e^{-i\Omega t}$ and reformulation of (\ref{eq-continuity-full})-(\ref{eq-Poisson-full}) as an operator eigenvalue problem:
\begin{equation}
    \label{eq-Matrix}
    -i(\hat{H}+\hat{V}_{drift}+\hat{V}_{sc}+\hat{V}_{visc})\mathbf{\Phi} = \Omega\mathbf{\Phi}.
\end{equation}
Above, we have introduced the ''two-component wave function'' $\mathbf\Phi = \{ n, u\}^{\rm T}$ describing density and velocity variations in plasma modes. The unperturbed motion is described by operator
\begin{equation}
    \label{eq-def-L}
    \hat{H} = 
    \begin{pmatrix} 
        0 & \partial_x[n_0(x)\cdot] \\
        \frac{e^2}{m^*}\int_0^L {\mathrm dx^\prime \partial_x G(x,x^\prime)\cdot} & 0
    \end{pmatrix}.
\end{equation}
The steady carrier drift, scattering, and viscosity act as small perturbations given by operators 
\begin{gather}
    \label{eq-def-Vdrift}
    \hat{V}_{drift} =\hat I \partial_x[v_0(x)\cdot]\\
    \label{eq-def-Vsc}
    \hat{V}_{sc} =\hat E \tau_{\rm p}^{-1};\qquad
    \hat{V}_{visc} =-\frac{\hat E}{m n_0}\partial_x\left[\eta \partial_x\cdot \right],
\end{gather}
here $I_{ij} = \delta_{ij}$ is the identity matrix, $\delta_{ij}$ is the Kronecker delta, and $E_{ij} = \delta_{ij}\delta_{i2}$.

At this stage, one might be willing to apply a standard Rayleigh-Schrodinger perturbation theory for corrections to eigen frequency $\delta \Omega_\lambda = \langle \Phi_\lambda | \hat V | \Phi_\lambda \rangle$, where $\lambda$ enumerates the plasmon modes. However, this step is premature until the inner product is specified. Apparently, a standard definition $\langle \Phi_\lambda | \Phi_{\lambda'} \rangle = \int dx[ n_\lambda^* n_{\lambda'} +u_\lambda^* u_{\lambda'}]$ fails: it neither ensures the hermiticity of dynamic matrix $\hat H$ nor even has a well-defined dimensionality. The 'proper' definition of inner product is inherited from the Lagrange function of a charged fluid:
\begin{multline}
    \langle {\Phi}_\lambda |\hat{\mathcal L}|{\Phi}_{\lambda'} \rangle \equiv e^2 \int_0^L{dx dx' n_{\lambda'}(x') G(x,x') n_\lambda (x)} - \\
    -    m \int_0^L {dx n_0 u_{\lambda'}(x) u_\lambda (x)}.
\end{multline}
With the above inner product, the dynamic matrix is hermitian under additional requirement $n(0) = n(L) = 0$. The latter boundary condition is physically equivalent to the fixation of carrier density by highly doped contacts.

The corrections to the eigen frequency of plasmon modes now acquire the form  
\begin{equation}
    \label{eq-pert-obvious}
    \delta\Omega_\lambda = -i\frac{\bra{\Phi_\lambda}\hat{\mathcal L}(\hat{V}_{drift}+\hat{V}_{sc}+\hat{V}_{visc})\ket{\Phi_\lambda}}{\bra{\Phi_\lambda}\hat{\mathcal L}\ket{\Phi_\lambda}}.
\end{equation}
Explicit evaluation of matrix elements in (\ref{eq-pert-obvious}) with the aid of current conservation and Bernoulli laws in the steady state allows us to present the correction to eigen frequency in a physically appealing form
\begin{equation}
    \label{eq-main-physical}
    \delta\Omega_\lambda = \frac{i}{\Pi}\left\{j_0 \left[K(L) - K(0) \right] - Q_{loss}\right\},
\end{equation}
where $K(x) = m u_\lambda(x)^2/2$ is the local kinetic energy in a plasmon mode,
\begin{equation}
    \Pi = e^2\int\limits_0^L{ \mathrm dx \mathrm dx^\prime n_\lambda(x) G(x,x^\prime) n_\lambda(x^\prime)}
\end{equation}
is the potential energy of interacting charge density fluctuations, and
\begin{equation}
  Q_{loss} = \frac{1}{2} \int\limits_0^L {\mathrm dx \left\{ \frac{e^2 n_0}{m\Omega^2 \tau_{\rm p}} E_\lambda^2 - u_\lambda \partial_x [\eta \partial_x u_\lambda ]\right\} }
\end{equation}
is the energy loss due to viscous friction and momentum non-conserving scattering.

Equation (\ref{eq-main-physical}) is the main point of our theory. We readily observe that the origin of plasmon growth, ${\rm Im}\,\delta\Omega>0$, is the excess of kinetic energy entering the mode at source over the energy lost at the drain. The energy constraints on turbulence in charge-neutral incompressible fluids date back to Reynolds~\cite{Reynolds1895} and Orr~\cite{Orr1907}, while energy balance approach was adapted to DS instability in Ref.~\onlinecite{cheremisin1999d} in the restrictive gradual-channel approximation. A closed-form expression for gain/loss in arbitrary nanostructure with 2DES (\ref{eq-main-physical}) appears for the first time. 

The necessity for asymmetry to achieve wave growth now becomes apparent. Indeed, for zero-order functions with certain parity $u^2(L) = u^2(0)$, and only the loss term retains in (\ref{eq-main-physical}). For modes without parity, the compensation of ingoing and outgoing energy can appear only accidentally. Therefore, the plasmon modes of any asymmetric nanostructure with perfect electron fluid ($\eta = 0$, $\tau^{-1}_{sc} = 0$) are unstable with respect to arbitrary weak drive current. This property hallmarks the instabilities of confined hydrodynamic plasmons from Cerenkov~\cite{mikhailov1998plasma} and beam~\cite{krasheninnikov1980instabilities} instabilities in extended systems. The latter typically develop above the threshold drift velocity order of plasmon phase velocity.

We stress that the density $n_\lambda(x)$ and velocity $u_\lambda(x)$ profiles in Eq.~(\ref{eq-main-physical}) are calculated in the absence of drift and dissipative effects, in line with meaning of perturbation theory. In other words, the excitability of a mode by current is fully determined by its ground state. At the same time, these zero-order profiles in an arbitrarily complex dielectric/gate environment can be obtained with commercial electromagnetic simulators. Already this fact makes our relation (\ref{eq-main-physical}) a valuable tool for design of high-gain plasmonic oscillators. In section III we are going to reveal reveal the conditions of large gain in common FETs with 2d channels, while in section IV we shall propose and substantiate asymmetric structures with ultrahigh gain.

However, large growth rate of instability would require large current which may be limited by heating, optical phonon scattering and flow choking~\cite{dyakonov1995choking} effects. Therefore, low threshold current is not less important than high gain. Fortunately, the developed perturbation theory also provides a route toward minimization of current,
\begin{equation}
\label{Variational-current}
    J^{th}\{ u \} = \frac{ \int\limits_0^L {\mathrm dx n_0(x) \left[ \tau^{-1}_{sc}u_\lambda^2 + \nu  |\partial_x  u_\lambda|^2 \right] }}{u^2(0) - u^2(L)} \rightarrow \min.
\end{equation}
The optimization in (\ref{Variational-current}) with respect to parameters of resonators is a routine numerical problem that will be discussed elsewhere. More importantly, equation (\ref{Variational-current}) allows us to set a lower bound on threshold current in an arbitrary plasmonic resonator. This bound will be derived in section V.

\section{Instabilities in common transistor structures}

Most general properties of current-driven hydrodynamic instabilities can be revealed on example of a 2d FET with closely located metal gate. In this case, the electric potential is bound to the local density via capacitance, $G(x,x^\prime)\approx (4\pi d/\varepsilon)\delta(x-x^\prime)$, and the potential energy of the mode becomes the energy of a non-uniformly charged capacitor
\begin{equation}
    \Pi = \frac{4\pi d e^2}{\varepsilon}\int_0^L{dx n^2_\lambda(x)}.
\end{equation}

\begin{figure}[t]
\includegraphics[width=\linewidth]{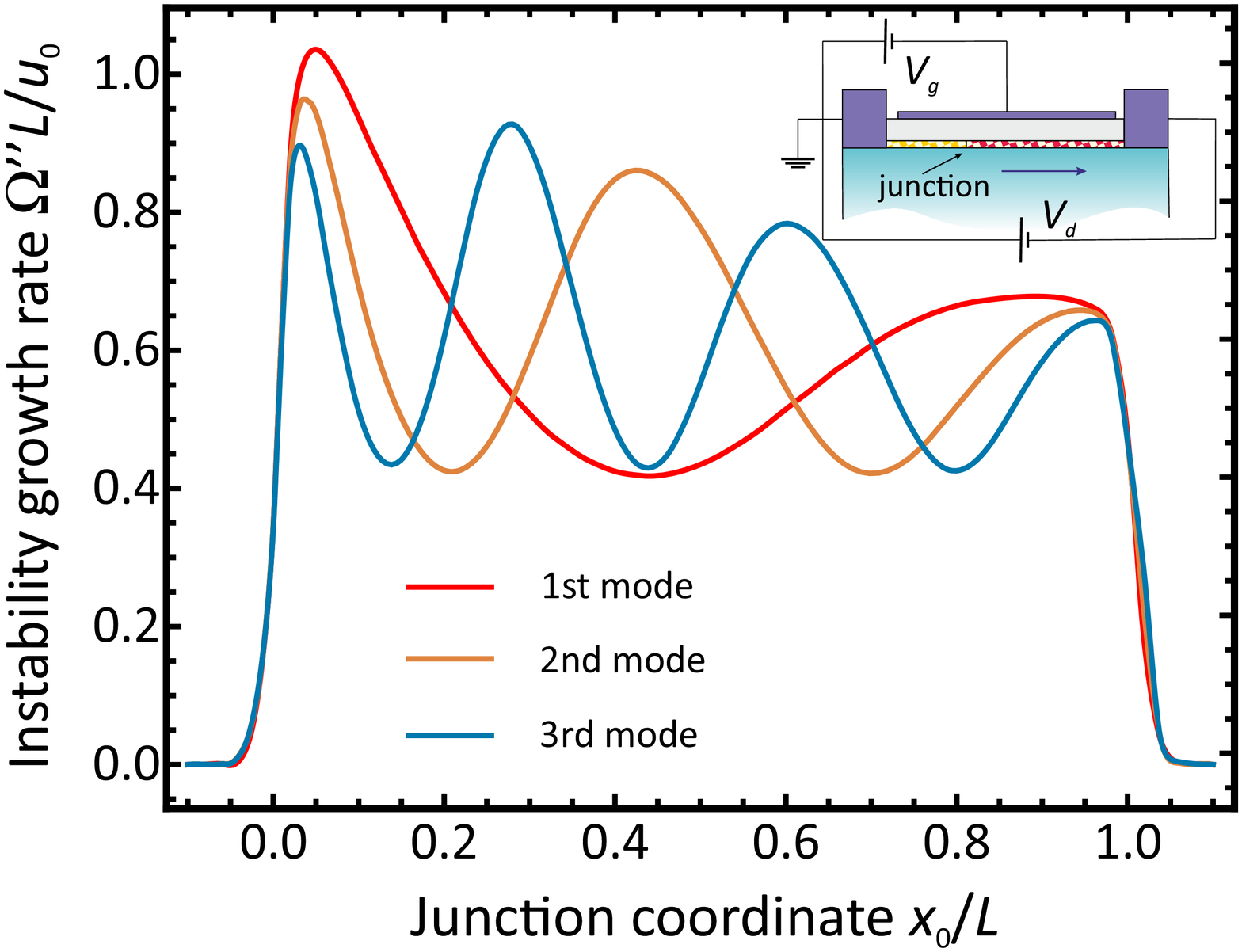}
\caption{\label{Fig1} 
Calculated growth rates of plasma instability in a gated 2DES (shown in the inset) with a steplike density profile on the location of the junction $x_0$. Growth rate is normalized by $u_0/L$, where $u_0$ is the drift velocity at the source. The distribution of carrier density is $n(x) = n_1 + (n_2 - n_1)[1+e^{(x-x_0)/l_j}]^{-1}$ with density contrast $n_1/n_2 = 0.58$, junction length $l_j = L/50$. The growth rate turns to zero for a symmetric structure with uniform carrier density (which corresponds to $x_0 < 0 $ or $x_0 > L$). The growth rates given by Eq. (\ref{eq-main-physical}) overlap with the ones obtained numerically.}
\end{figure}

The mode asymmetry in a gated FET is most simply achieved by introducing a carrier density step ($n^+-n$ junction) in the channel (e.g., by non-uniform doping or split gate). Such model system allows an analytical solution for the instability growth rate that perfectly matches our perturbative formula (\ref{eq-main-physical}) [see Supporting section I]. The calculated growth rates in such a setup are depicted in Fig.~\ref{Fig1} for a perfect electron fluid as a function of the junction position for the three lowest eigenmodes. First of all, the oscillations grow in time if electron drift is directed from a low-density to a high-density region. This can be understood with current continuity equation: high density at the drain implies weak oscillations of carrier drift velocity and small loss of kinetic energy at the drain side. Apart from coordinate-independent part, the growth rate demonstrates position-dependent fringes with period order of plasmon wavelength.

When the contrast of densities at the $n^+-n$ junction is high, the plasmon field is confined to a highly doped region. Such an abrupt junction effectively mimics the short-circuited drain from an original proposal by Dyakonov and Shur, which can be confirmed by integration of continuity equation across the boundary. For abrupt junction in the channel of a gated 2DES, our perturbative analysis yields the growth rate $\delta \Omega \simeq i u_0/L_{n^+}$, where $L_{n^+}$ is the length of highly doped section, in agreement with the original result of DS.

\begin{figure}[t]
\includegraphics[width=\linewidth]{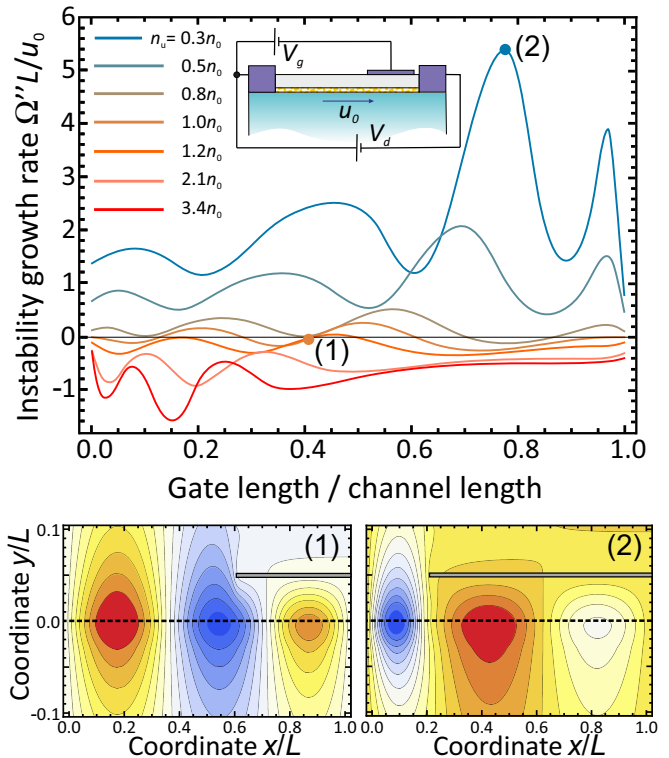}
\caption{\label{Fig2} 
Top: calculated instability growth rates for the third plasmon mode in a partly gated FET (shown in inset) with different gate lengths and carrier densities. The growth rates are normalized by $u_0/L$, where $u_0$ is the drift velocity at the drain (same for all curves), the density at the drain $n_0$ is also fixed. The instability benefits if the drift is directed from low- to high-density region and is especially pronounced if low-density region is ungated. A structure with uniform density (orange line) also supports instabilities due to the asymmetry of electrical environment.
Bottom: spatial distribution of plasmon potential in partly gated FET for two modes: (1) with almost symmetric in-plane distribution of potential and zero gain (2) with highly asymmetric in-plane potential and high gain.}
\end{figure}

Formation of a density step is most easily achieved with a metal gate partially covering the channel, as it was done in Ref.~[\onlinecite{ElFatimy2010algan}]. This gate already introduces structural asymmetry translating in the asymmetry of plasmonic modes, even in the absence of gate voltage. In Fig.~\ref{Fig2}, orange line, we show the growth rates of plasmon modes in FET with {\it uniform} carrier density and metal gate adjacent to drain side [calculated using perturbative formula (\ref{eq-main-physical}) and spectral method for unperturbed problem, see Supporting section II]. It demonstrates an oscillatory behaviour with varying gate length with a relatively small maximum growth rate $\sim u_0/L$, where $u_0$ is the drift velocity in the ungated region. Addition of extra density asymmetry, $n_g \neq n_u$, results in much more pronounced instability (for $n_g > n_u$) or more pronounced current-induced plasmon decay (for $n_g < n_u$). The rule for maximization of growth rate is here the same as for fully gated FET: electron drift should be directed from low-density toward high-density region. This observation is in qualitative agreement with experimental results~\cite{ElFatimy2010algan}, where the depletion of a region near the source reduced the threshold voltage for negative differential resistance and THz emission.



\section{Plasmonic oscillators with high gain}

The modes mostly subjected to current-driven instabilities are those having very large electric field at the source and vanishing field at the drain. This leads to a large difference between incoming and outgoing energy fluxes. Unfortunately, an enhancement of current-driven growth rate for highly asymmetric modes comes at the cost of elevated viscous dissipation due to non-uniformity of drift velocity.

\begin{figure}[t]
\includegraphics[width=0.9\linewidth]{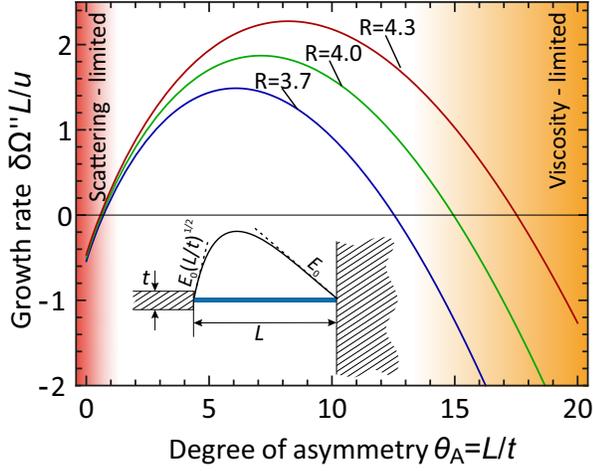}
\caption{\label{Fig3} 
Schematic dependence of instability growth rate on the degree of asymmetry in plasmonic mode $\theta_A$. For nearly symmetric mode profiles, the gain is overwhelmed by momentum non-conserving scattering (red-shaded area). For strongly asymmetric modes, the instability is stabilized by viscosity (orange-shaded area). For intermediate asymmetry, the growth rate reaches a maximum exceeding $u_0/L$. The curve is plotted with eq.~(\ref{Growth-keen-blunt}) derived for a 2DES with keen source and blunt drain contacts (shown in inset). The ratio of channel length $L$ to the source thickness $t$ characterizes the degree of asymmetry $\theta_A = L/t$
}
\end{figure}

The simplest setup illustrating gain-viscosity compromise represents a 2d channel with keen source and blunt drain contacts, the latter will be modelled as an infinite conducting wall. The degree of asymmetry in such a setup can be characterized by the ratio of channel length $L$ to the source thickness $t$ which we assume to be large, $\theta_A = L/t \gg 1$.  The ratio of electric fields near the source electrode and drain will be order of $\theta_A^{1/2}$, due to a square-root singularity of electric field of a wedge in two dimensions. Therefore, the current-driven energy gain would scale linearly with degree of asymmetry $\theta_A$. The strong field will span over the length $\delta L \simeq L /\theta_A^{1/2}$ from the drain. This allows us to find the scaling of energy dissipation due to viscosity:
\begin{equation}
    Q_{visc} \propto \nu |\partial_x E|^2 \delta L \propto \frac{\nu}{L^2} \theta_A^{3/2}.
\end{equation}
What concerns scattering-induced dissipation, it leads to a trivial downshift of instability growth rate by $1/2\tau_{\rm p}$, independent of mode profile. Combining the above estimates, we deduce the scaling of current-driven growth rate with degree of asymmetry $\theta_A$:
\begin{equation}
\label{Growth-keen-blunt}
    \delta\Omega'' \simeq \frac{u_0}{L} \theta_A - \frac{\nu}{L^2} \theta_A^{3/2} - \frac{1}{2\tau_{\rm p}},
\end{equation}
where we have omitted numerical prefactors order of unity in terms with $\theta_A$, and used the fact that resonant plasma frequency $\omega_p \simeq (n_0 e^2/mL)^{1/2}$ [\onlinecite{Ryzhii2004slot}]. The resulting growth rate has a maximum achieved at
\begin{equation}
    \theta_A^* = \frac{4}{9} \left(\frac{u_0}{\nu L}\right)^2,
\end{equation}
the term in the round brackets is nothing but the Reynolds number ${\rm R} = u_0/\nu L$. In the setup with optimized asymmetry, the growth rate of plasmon mode will be order of
\begin{equation}
    \delta\Omega''_{\max} \simeq \frac{4}{27}\frac{u_0}{L} {\rm R}^2 - \frac{1}{2\tau_{\rm p}},
\end{equation}
i.e. $4R^2/27$ times higher than in an original proposal of Dyakonov and Shur.


\section{Lower bound on the instability threshold current}

Thus far, we have demonstrated the maximization of instability growth rate using Eq.~(\ref{eq-main-physical}) with respect to the geometrical parameters of plasmonic resonator. It is possible to go even further and find the lower bound of threshold current with respect to {\it all possible plasmonic modes}, or, equivalently, all possible geometries of plasmonic resonators. Already from scaling considerations it is apparent that in the absence of scattering this minimum would be order of $\nu n_0/L$.

The lower bound on threshold current is most easily found for uniformly doped channel. We map the two-dimensional channel to the region $\xi \in [0;1]$ and rewrite the velocity through electric potential. This results in the functional
\begin{equation}
\label{Variational-potential}
    J^{th}\{\varphi\} = \frac{n_0 \nu}{L}\frac{\int_0^1{d\xi [\varphi''^2(\xi) +p^2 \varphi'^2(\xi)]} }{\varphi'^2(1) - \varphi'^2(0)},
\end{equation}
where $p^2 = L^2/\tau_{\rm p}\nu$ is the dimensionless parameter showing the relative role of scattering and viscous dissipation. Minimizing (\ref{Variational-potential}) one should imply fixed-potential boundary conditions $\varphi(0) = \varphi(1) = 0$.

\begin{figure}[t]
\includegraphics[width=0.9\linewidth]{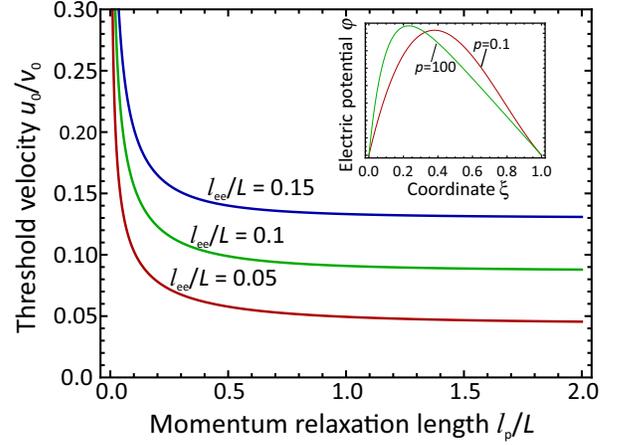}
\caption{\label{Fig4} 
Dependence of {\it minimal} threshold velocity for the onset of instability $u_0$ (in units of carrier velocity $v_0$) on momentum relaxation length $l_p$ (in units of channel length $L$). Curves are plotted for various mean free path with respect to e-e collisions $l_{\rm ee}/L$. Inset: distribution of electric potential in the plasmonic mode with ''optimal'' asymmetry corresponding to weak momentum non-conserving scattering ($p=0.1$) and strongly diffusive transport ($p=100$)}
\end{figure}

The minimization is quite lengthy yet possible (Supporting section III), and results in a nice expression for the minimum threshold current
\begin{equation}
\label{Eq-critical-current}
    J^{th}_{\min} =\frac{\nu n_0}{L} \frac{p}{2}\frac{\sqrt{F(p)^2-1}}{F(p) - F^2(p/2)},
\end{equation}
where $F(p) = \sinh p/p$. In the absence of scattering ($p=0$), the minimum threshold current equals $2\sqrt{3} \nu n_0/L$. In other words, there exists a minimal Reynolds number in two-dimensional systems below which the instability is impossible:
\begin{equation}
\label{Reynols-crit}
    {\rm R}_{\min} = 2\sqrt{3}.
\end{equation}
With increasing the frequency of momentum non-conserving scattering, the relative role of viscosity is lowered, and the modes providing the minimal current become highly asymmetric. The threshold current apparently grows, and in the diffusive limit $p\gg 1$ is becomes
\begin{equation}
    J^{th}_{\min}  = \frac{\nu n_0 p}{2 L} \approx \frac{v_0 n_0}{4}\sqrt{\frac{\tau_{ee}}{\tau_{\rm p}}},
\end{equation}
where $v_0$ is the characteristic carrier velocity (thermal or Fermi), and we have used the fact that $\nu \approx v_0^2 \tau_{ee}/4$.

The critical Reynolds number in 2d electron fluid equal to $2\sqrt{3}\approx 3.5$ is quite low compared to typical Reynolds numbers in fluid turbulence $\sim 10^3$. Therefore, it is achievable in current solid-state experiments. Indeed, we can rewrite Eq.~(\ref{Reynols-crit}) in terms of drift velocity as
\begin{equation}
    \frac{u_0}{v_0} \ge \frac{\sqrt{3}}{2}\frac{l_{ee}}{L}.
\end{equation}
For a typical graphene-based transistor ($L=1$ $\mu$m) at room temperature $l_{ee}\approx 200$ nm~\cite{kumar2017superballistic}, and the critical velocity is just $0.17$ of the Fermi velocity. If the mobility is realistically high, $\mu \approx 5\times 10^4$ cm$^2$/(V s), the momentum non-conserving scattering length is order of $v_0 \tau_{\rm p} \approx 1$ $\mu$m, and the scattering parameter is estimated as $p \approx 4.5$. Such scattering, according to (\ref{Eq-critical-current}), raises the critical velocity just by $15$ \%.

The above estimates are strict {\it lower bounds} for threshold velocity leading to instability, and they are reached for some ''optimally asymmetric'' modes. The modes supported by real asymmetric devices can be different from this optimum. However, one should keep in mind that our minimal current was obtained under assumption of uniform doping and fixed potential and density at the contacts; relaxation of these assumptions may change the lower bound.



\section{Discussion and conclusions}

To conclude, we have revealed the link between the (a)symmetry of confined structures with two-dimensional electron fluid and the (in)stability of their plasmonic modes under direct current. Namely, we have shown that plasmon gain under direct current is an inherent property of any asymmetric plasmonic nanostructure, independent on the origin of asymmetry. The latter can be caused by non-uniformity of dielectric environment, non-uniform distribution of carrier density, or non-equivalence of source and drain contacts. The well-known Dyakonov-Shur instability occurring in FETs with grounded source and short-circuited drain~\cite{dyakonov1993shallow} appears just as a particular case of asymmetry-driven instabilities.

We have provided a constitutive relation between current-induced plasmon gain, viscosity- and scattering induced loss, and the distributions of plasmon potential $\varphi_0(x)$ in the absence of drift and dissipative effects. This expression is based on perturbation theory that we have specially developed for the hydrodynamics of two-dimensional charged fluid. With this relation, it becomes possible to study the 2d plasmon instabilities without going into complicated simulations of turbulent flows~\cite{Koseki2016giant}. All necessary ingredients to judge on the possibility of gain are the plasmon mode profiles $\varphi_0(x)$ (''zero-order functions'') that can be obtained with commercial electromagnetic simulators. Our perturbative expressions for plasmon eigen frequency allowed to reveal the competition between current-driven gain and viscosity both of which increase with the degree of asymmetry. Moreover, they allowed us to find an ''optimal degree of asymmetry'' for which the plasmon gain is highest or instability threshold current is the lowest. There exists a universal lower bound of Reynolds number for development of plasmon instability in 2DES with fixed voltage drop between source and drain, which equals ${\rm R}_{\min} = 2\sqrt{3}$.

Despite an apparent universality of the developed perturbation theory for hydrodynamic plasmons in 2DES, there remains a plenty of directions to which the theory can be extended. So far we have considered one-dimensional collective oscillations which are the only active in FETs with relatively narrow channels of width $W \ll L$. For wider channels, edge modes propagating along the contact terminals or gated/ungated boundaries can be also excited by dc current~\cite{dyakonov2008boundary}. An extension of theory for two-dimensional oscillations faces the potential problem of edge mode decay into bulk even in the absence of dissipation~\cite{Mikhailov1992interedge}. In other words, such theory would deal with inherently quasi-stationary states, which is a complicated yet doable problem~\cite{Weiss2016pertopencavity}.

The perturbation theory is readily extended to periodic 2DES -- plasmonic crystals~\cite{dyer2013induced}. It is possible to show that plasmon gain linear in drive current does not appear in such structures~\cite{petrov2017amplified,kachorovskii2012current}. The first-order correction to plasmon frequency induced by current in gated crystal with modulated doping reads
\begin{equation}
    \frac{\delta \Omega_\lambda}{\Omega_{0\lambda}} =\frac{1}{\Pi}\int_0^L {\mathrm dx u_0 {\rm Re}n_\lambda u^*_\lambda},
\end{equation}
where $\Omega_{0\lambda}$ is the mode frequency in the absence of dc current, and the integration is performed over the unit cell. The correction is entirely real and represents nothing but current-induced Doppler shift. The absence of gain here can be explained by equality of energy fluxes going in and getting out of each cell, by the virtue of crystal periodicity. However, plasmon gain can appear in the second order of perturbation theory.

So far our treatment has been limited to hydrodynamic regime of electron transport realized at frequencies $\omega\tau_{ee} \ll 1$. It is tempting to generalize the relation of instability and asymmetry also to the collisionless (kinetic) regime. The concept of amplified reflection as origin of instability~\cite{sydoruk2012amplifying} in bounded 2DES hints that hydrodynamic transport is not a necessary condition of instability~\cite{Dmitriev2001collisionless}.

Finally, our theory of instabilities was developed for bounded 2DES with fixed voltage drop between terminals. This is a seemingly good approximation, but in reality all metal contacts have very large inductive impedance at microwave frequencies. With inclusion of this effect, the plasma oscillations start interacting with electrical oscillations in $LC$ circuit formed by contact inductance and source- and drain-to-gate capacitance. Extra origins of gain due to back-action of induced currents in contact pads on plasma wave look possible in this regime~\cite{ryzhii2005transit}.   

The work was supported by the grant \# 16-19-10557 of the Russian Science Foundation. The authors are grateful to D. Bandurin for providing Refs.~\onlinecite{Kopylov1980instability,kopylov1987perioddoubling} and to Zh. Devizorova and M.S. Shur for helpful discussions.

\section*{Supporting Information: numerical method}

In order to obtain Figs. \ref{Fig1} and \ref{Fig2} we applied a standard spectral numerical method to the system of hydrodynamic equations (\ref{eq-continuity-full}-\ref{eq-Poisson-full}) with Chebyshev polynomials of the first kind $T_i$ taken as the basis functions. To be more concrete, after writing the linearized Eqs. (\ref{eq-continuity-full}), (\ref{eq-Euler-full}) in dimensionless units ($x\rightarrow \xi=2x/L - 1$, $n\rightarrow \delta n = n/n_0(0)$, $u \rightarrow \delta u/s(0)$, $\Omega\rightarrow\omega = \Omega L/s(0)$ where $s(0)^2 = e^2n_0(0)L/m^*$) we substituted the Chebyshev expansions in the form $\left\{\delta n, \delta u\right\} = \sum_{i=0}^N C_i^{\delta n, \delta u}T_i(\xi)$ and projected the system on each of the polynomials $T_i(\xi),i = 0..N$. After these manipulations we come to:
\begin{equation}
\label{eq-numricalSystem}
    \begin{pmatrix} 
        \hat{M}^{(1)} & \hat{M}^{(2)} \\
        \hat{M}^{(3)} & \hat{M}^{(4)}
    \end{pmatrix}
    \begin{pmatrix} 
        C^{\delta n}_i \\
        C^{\delta v}_i
    \end{pmatrix} = \mathrm i\omega
    \begin{pmatrix} 
        C^{\delta n}_i \\
        C^{\delta v}_i
    \end{pmatrix},
\end{equation}
where 
$$\hat{M}_{ij}^{(1)} = \hat{M}_{ij}^{(4)} = t_{ij}\bra{T_j}w(\xi)\partial_\xi \ket{v_0 T_i},$$
$$\hat{M}_{ij}^{(2)} = t_{ij}\bra{T_j}w(\xi)\partial_\xi\ket{n_0 T_i},$$ 
$$\hat{M}_{ij}^{(3)} = t_{ij}\bra{T_j}w(\xi)\partial_\xi\ket{\int_{-1}^1 d\xi'G(\xi,\xi')T_i(\xi')},$$
$$ t_{ij} = \begin{cases} 1/\pi, i=0, \\ 2/\pi, \text{otherwise} \end{cases},$$
and $w(\xi) = (1-\xi^2)^{-1/2}$ is the weight function; $i=0..N, j=0..N$ for all the matrices. Now we shall imply boundary conditions which require $\delta n(-1)=\delta n(1)=0$; that leads to $C_N^{\delta n}=-\sum_{i=0}^{N/2-1} C_{2i}^{\delta n};$ $C_{N-1}^{\delta n}=-\sum_{i=0}^{N/2-1} C_{2i+1}^{\delta n}$ ($N$ is supposed to be even). We use these expressions to eliminate $C_N^{\delta n}$ and $C_{N-1}^{\delta n}$ from the system (\ref{eq-numricalSystem}) and, in order to keep the matrix dimensions, truncate the first three matrices. We get:
$$\hat{M}_{ij}^{(1)} =\hat{M}_{ij}^{(1)} - \hat{M}_{i\{N,N-1\}}^{(1)}, i= 0..N-2, \,j=0..N-2;$$
$$\hat{M}_{ij}^{(2)} = \hat{M}_{ij}^{(2)}, i= 0..N, \,j=0..N-2;$$
$$\hat{M}_{ij}^{(3)} = \hat{M}_{ij}^{(3)} - \hat{M}_{i\{N,N-1\}}^{(3)}, i= 0..N, \,j=0..N-2,$$
where the notation $\{N,N-1\}$ means that we take $N$ if $j$ is even and $N-1$ otherwise.

It is worth mentioning that one may face computational difficulties while evaluating matrix elements $\hat{M}_{ij}^{(3)}$ as the Green's function is singular on the diagonal $\xi=\xi'$. To overcome this, we extracted the singular part of the Green's function:
$$G = (G-G_{sing})+G_{sing} = G_{reg}+G_{sing},$$
where $G_{sing}=\ln\dfrac{(\xi-\xi')^2}{(\xi+\xi'-2)^2(\xi+\xi'+2)^2}$ accounts for the singularity provided by charge itself as well as by the two nearest mirror images in the electrodes. The regular integral was then calculated numerically while the singular one can be significantly simplified via transition in the complex plane. 

The values of the first correction obtained by the described procedure and by the perturbation theory (Eq.(\ref{eq-main-physical})) totally coincide at small drift velocities.

\section*{Supporting information: Instability in gated 2DES with carrier density step}
Within this section, we shortly discuss the derivation of the dispersion relation for plasma waves in a fully gated FET with an $n^+-n$ junction. The derivation incorporates a common\cite{kachorovskii2012current,petrov2016plasma,petrov2017amplified} decomposition of a plasma mode into two modes (we denote them $n^+$ and $n$ as they are confined to each of the regions) which can be presented in the form
\begin{equation}
    \zeta^{(n,n^+)} = \zeta^{(n,n^+)}_\rightarrow \exp\left(ik_\rightarrow^{(n,n^+)}x \right) + \zeta^{(n,n^+)}_\leftarrow \exp\left(ik_\leftarrow^{(n,n^+)}x \right),
\end{equation}
where $\zeta$ stands for some physical property of the wave (potential, velocity, etc.), $\zeta_\leftrightarrow$ are the amplitudes of forward and backward travelling waves and $k_\leftrightarrow = \Omega/u\pm s$ denote their wavevectors\cite{dyakonov1993shallow}, $u$ and $s$ are the drift velocity and plasma wave velocity (apparently, they differ for each of the regions).

These modes are connected via boundary conditions at the $n^+-n$ interface, which we take to be the ''natural'' boundary conditions of current and energy flow conservation, following from the initial Eqs. (\ref{eq-continuity-full})-(\ref{eq-Euler-full}). Keeping in mind the demand for zero ac potential at the contacts, we obtain four equations to determine four unknown amplitudes. The determinant of this system should equal zero which allows to obtain the dispersion law:
\begin{widetext}
\begin{multline}
    (s_n-s_{n^+})\left[(u_{n^+}s_n+u_ns_{n^+})\cos(\varphi_n-\varphi_{n^+})+i(u_n u_{n^+}+s_ns_{n^+})\sin(\varphi_n-\varphi_{n^+})\right]+\\
    (s_n+s_{n^+})\left[(u_{n^+}s_n-u_ns_{n^+})\cos(\varphi_n+\varphi_{n^+})+i(u_n u_{n^+}-s_ns_{n^+})\sin(\varphi_n+\varphi_{n^+})\right] = 0,
\end{multline}
\end{widetext}
where $\varphi_{n,n^+} = (k^{n,n^+}_\rightarrow - k^{n,n^+}_\leftarrow)L_{n,n^+}/2.$

\section*{Supporting Information: solution of the variational problem}
In this section, we solve the variational problem (\ref{Variational-potential}) and find the ''optimal'' plasmon mode which is excited by the smallest direct current. To this end, we find the variation $\delta J^{th} = J^{th}\{\varphi_0+\delta \varphi\} - J^{th}\{\varphi_0\} $ and set it to zero
\begin{multline}
 \delta {{J}^{\text{th}}}=\int\limits_{0}^{1}{\left[ \varphi_0^{IV}\left( x \right)-{p^2}{\varphi''_0}(x) \right]\delta \varphi ( x )dx}+\\
 \left. \left[ {\varphi''_{0}} ( x )-{{J}^{\text{th}}}{\varphi'_{0}}( x ) \right]\delta {\varphi }'\left( x \right) \right|_{0}^{1}-\\
 \left. \varphi'''_{0}( x )\delta \varphi \left( x \right) \right|_{0}^{1}=0.  
\end{multline}

The first term provides us with the differential equation for optimal mode 
\begin{equation}
\label{Eq-var}
 \varphi^{IV}_0 (x) - p^2 \varphi_0''(x)=0.
\end{equation}
The second term yields two boundary conditions (BCs)
\begin{gather}
\label{Eq-var-bc1}
    {\varphi''_{0}} ( 0 )-J^{\text{th}}\{\varphi_{0}\}{\varphi'_{0}}( 0 ),\\
    {\varphi''_{0}} ( 1 )-J^{\text{th}}\{\varphi_{0}\}{\varphi'_{0}}( 1 ).
\end{gather}
The last term vanishes identically due to fixation of electric potential at the contacts, which yields two remaining boundary conditions
\begin{equation}
\label{Eq-var-bc2}
    \varphi_0(0) = \varphi_0(1) = 0.
\end{equation}
The solution of (\ref{Eq-var}) with BCs (\ref{Eq-var-bc2}) is
\begin{multline}
\label{Var-solution}
  \varphi_0(x) =  A \left(-e^p x+e^{p x}+x-1\right)-\\
  B(e^{-p} x - e^{-p x} - x + 1).
\end{multline}
Another set of BCs (\ref{Eq-var-bc2}) couples the constants $A$ and $B$ to the value of functional on the sought-for function $J^{\text{th}}\{\varphi_{0}\}$. We shall not use them but take a simpler way. Without loss of generality we can set $B=1$ in (\ref{Var-solution}), and evaluate $J^{\text{th}}\{\varphi_{0}\} \equiv J^{\text{th}}(A)$:
\begin{equation}
   J^{\text{th}}(A) = \frac{\nu n_0 p^2}{2 L} \frac{ e^{2 p} \left[A^2 \left(e^{2 p}-1\right)+4 A p+1\right]-1}{[e^p-1] [e^p (p-2)+p+2] [A^2 e^{2 p}-1]}.
\end{equation}
Minimization of $|J^{\rm th}|$ with respect to $A$ provides two optima
\begin{equation}
\label{Variational-parameter}
    A_{\pm} = - e^{-p} \left( \frac{\sinh p}{p} \pm \sqrt{\frac{\sinh^2 p}{p^2} - 1}\right),
\end{equation}
they correspond to opposite currents of equal magnitude given by Eq.~(\ref{Eq-critical-current}). It is easy to check that boundary conditions (\ref{Eq-var-bc1}) with $A_\pm$ from above equation are satisfied automatically.

\bibliography{refs}

\end{document}